\documentclass[11pt]{article}
\usepackage{amsmath,amssymb,graphicx}
\usepackage{hyperref}
\usepackage{cite}

\usepackage{multicol,color,longtable}

\definecolor{darkred}{rgb}{0.5,0,0.5}
\hypersetup{pdfborder={0 0 0},colorlinks=true,urlcolor=darkred,citecolor=blue,linkcolor=darkred}

\makeatletter

\@addtoreset{equation}{section}
\makeatother

\newcommand{\nn}{\nonumber}

\newcommand{\ints}{\mathbb{Z}}
\newcommand{\reals}{\mathbb{R}}

\newcommand{\lb}{\left[}
\newcommand{\rb}{\right]}
\newcommand{\ld}{\left\{}
\newcommand{\rd}{\right\}}
\newcommand{\lp}{\left(}
\newcommand{\rp}{\right)}
\newcommand{\id}{1\!\!1}

\newcommand{\eps}{\epsilon}

\newcommand{\qchi}{\hat{\chi}}
\newcommand{\qpsi}{\hat{\psi}}
\newcommand{\cA}{\mathcal{A}}
\newcommand{\cB}{\mathcal{B}}
\newcommand{\cC}{\mathcal{C}}
\newcommand{\cD}{\mathcal{D}}

\newcommand{\ta}{{\tt{a}}}
\newcommand{\tb}{{\tt{b}}}
\newcommand{\tc}{{\tt{c}}}
\newcommand{\td}{{\tt{d}}}
\newcommand{\te}{{\tt{e}}}
\newcommand{\tf}{{\tt{f}}}

\newcommand{\qphi}{\hat{\phi}}
\newcommand{\hJ}{\hat{J}}
\newcommand{\hA}{\hat{{\bf A}}}
\newcommand{\hB}{\hat{{\bf B}}}

\newcommand{\ad}[1]{\mathrm{ad}\,#1}

\newcommand{\mf}[1]{{\mathfrak{#1}}}
\newcommand{\emp}[1]{{\em #1}}

\begin{document}

\thispagestyle{empty}

{\flushright {\tt AEI-2013-220}\\[20mm]}

\begin{center}
{\LARGE \bf \sc On higher spin realizations of $K(E_{10})$}\\[10mm]

\vspace{8mm}
\normalsize
{\large  Axel Kleinschmidt${}^{1,2}$ and Hermann Nicolai${}^1$}

\vspace{10mm}
${}^1${\it Max-Planck-Institut f\"{u}r Gravitationsphysik (Albert-Einstein-Institut)\\
Am M\"{u}hlenberg 1, DE-14476 Potsdam, Germany}
\vskip 1 em
${}^2${\it International Solvay Institutes\\
ULB-Campus Plaine CP231, BE-1050 Brussels, Belgium}

\vspace{15mm}

\hrule
\vspace{5mm}
\begin{tabular}{p{12cm}}
{\footnotesize Starting from the known unfaithful spinorial representations of the 
compact subalgebra $K(E_{10})$ of the split real hyperbolic Kac--Moody algebra
$E_{10}$ we construct new fermionic `higher spin' representations of this algebra (for 
`spin-$\frac52$' and `spin-$\frac72$', respectively) in a second quantized framework. 
Our construction is based on a simplified realization of $K(E_{10})$ on the Dirac and
the vector spinor representations in terms of the associated roots, and on a re-definition
of the vector spinor first introduced in \cite{Damour:2009zc}. The latter replaces 
manifestly $SO(10)$ covariant expressions by new expressions that are covariant 
w.r.t. $SO(1,9)$, the invariance group of the DeWitt metric restricted to the space
of scale factors. We present explicit expressions for all $K(E_{10})$ elements that
are associated to real roots of the hyperbolic algebra (of which there are infinitely many), 
as well as novel explicit realizations of the generators associated to imaginary roots 
and their multiplicities. We also discuss the resulting realizations of the Weyl group. }
\end{tabular}
\vspace{5mm}
\hrule

\end{center}

\setcounter{page}{0}
\newpage

\section{Introduction}

Starting with the discovery of the hidden $E_{7}$ symmetry of $N=8$ supergravity in $D=4$~\cite{Cremmer:1978ds}, the study of hidden symmetries has been one of the cornerstones in investigations of the landscape of supergravities. For maximal supergravity in $D\geq 2$ space-time dimensions, the hidden symmetry group is $E_{11-D}$ (in split real form)~\cite{Julia,Ni87a}. This fact has also been crucial in string theory where the associated discrete U-duality~\cite{Hull:1994ys} is instrumental for the construction of orbits of BPS states (for a review see~\cite{Obers:1998fb}) and for constraining non-perturbative effects~\cite{Green:1997tv,Pioline:1998mn,Green:2010wi,Green:2010kv,Fleig:2012xa}. It has also been conjectured that the infinite-dimensional Kac--Moody groups $E_{10}$ and $E_{11}$ play a central role for understanding dimensionally reduced supergravity~\cite{Julia} or even the uncompactified theory~\cite{West:2001as,Damour:2002cu}. All these proposals rest on a non-linearly realised symmetry $E_{11-D}$; the bosonic fields, both propagating and auxiliary, are associated with the coset $E_{11-D}/K(E_{11-D})$ where $K(E_{11-D})$ denotes the maximal compact subgroup of $E_{11-D}$. In this paper, we will restrict to $E_{10}$ whose Lie algebra will be defined below together with that of $K(E_{10})$.

As supergravity is a supersymmetric theory it is necessary to also include the 
fermionic degrees of freedom in the context of the $E_{10}$ proposal. Since the compact subgroup $K(E_{10})$ generalises the spatial Lorentz group $Spin(10)$ under which  the fermionic degrees of freedom transform, one needs to understand its representation theory. Even for the affine 
subgroup $E_9$ and its compact subgroup $K(E_9)$ this is a 
non-trivial problem \cite{NS}. First results for $K(E_{10})$  have been obtained 
in~\cite{deBuyl:2005zy,Damour:2005zs,deBuyl:2005mt,KN,Damour:2006xu,Kleinschmidt:2007zd,Damour:2009zc}, where it was shown how to write the supersymmetry 
parameter and the gravitino field as consistent albeit unfaithful representations of the 
Lie algebra of $K(E_{10})$.\footnote{We will (ab)use $K(E_{10})$ to denote both the group and Lie algebra.} The qualifier `unfaithful' here signals that~---~although the Lie algebra $K(E_{10})$ is infinite-dimensional~---~the representation spaces are finite-dimensional. The corresponding transformation rules have been worked out for maximal supergravity (and also for
half-maximal supergravity~\cite{Kleinschmidt:2004dy}); they have reappeared 
in recent studies of fermions in the context of generalised geometry, see e.g.
~\cite{Jeon:2011vx,Hohm:2011nu,Coimbra:2012af}. 
In the context of $E_{11}$, the essential parts of the supersymmetry parameter representation were identified in~\cite{West:2003fc} and the vector spinor given in~\cite{KN}. 
Dirac-type representations for other 
simply-laced Kac--Moody groups were recently considered in~\cite{Koehl}.

All representations of $K(E_{10})$ that have been constructed so far are restricted to 
the consideration of the fermionic variables which appear in supergravity, namely the 
gravitino vector spinor, and the Dirac spinor associated with the supersymmetry
parameter. However, a  systematic understanding of the $K(E_{10})$ representation 
theory has eluded all efforts so far. Even the less ambitious goal of identifying 
a {\em faithful} fermionic representation, or even only the construction of new unfaithful 
representations that are genuinely different from the ones having a direct origin 
in supergravity has remained out of reach so far (see also \cite{NS} for related
attempts for  $K(E_9)$). At least with regard to the latter problem
the present work overcomes the barrier by constructing 
fermionic `higher spin' representations of the $K(E_{10})$ algebra, associated to 
`spin-$\frac52$' and to `spin-$\frac72$', respectively.\footnote{This terminology 
 should not be taken too literally. For instance, when talking about 
 `spin-$\frac52$' we really mean a fermion operator that is {\em analogous} to the
 corresponding fermion field in four space-time dimensions. Note also that
 `spin' in $D=11$ dimensions is really an $SO(9)$ or $SO(10)$ (for massive
  excitations) representation label consisting of a 4-tuple or 5-tuple of 
  (half) integer numbers designating a representation of the little group.} The new spinorial representations of $K(E_{10})$ obtained in this way appear to lie `beyond supergravity', 
and thus are not expected to be derivable from any known model of supergravity.
Their very existence points to the existence of an infinite hierarchy of higher spin 
realizations of $K(E_{10})$, which become less and less unfaithful with increasing spin. 
At the very least, they should provide new tools to study  the algebra  of $K(E_{10})$. 
One useful tool in our investigation will be the truncation to better-known 
subalgebras of $K(E_{10})$, most notably $K(E_9)$ and $K(E_8)=D_8\equiv \mf{so}(16)$ 
that can be used to understand some of the features of our novel representations.
Indeed, the proof that the new representations are really inequivalent rests on 
such a decomposition.

Among other results, our work rests on recent results from investigations 
of fermionic cosmological billiards~\cite{Damour:2009zc} 
and quantum cosmology~\cite{Kleinschmidt:2009cv,Kleinschmidt:2009hv,Damour:2011yk,Damour:2013eua}. The main new insight (in particular inspired by the work 
of \cite{Damour:2009zc,Damour:2013eua}) is that by breaking spatial $SO(10)$ Lorentz covariance, simpler expressions for the $K(E_{10})$ transformations emerge. The breaking appears naturally in cosmological billiards where the effective bosonic degrees of freedom in mini-super-space are written more naturally in terms of $SO(1,9)$ transformations~\cite{Damour:2000wm}. The associated indefiniteness stems from 
the conformal mode of gravity that  appears with the `wrong' sign in Hamiltonian 
treatments of gravity, and is directly connected to the indefiniteness of the DeWitt 
metric on the Wheeler-DeWitt superspace of 3-geometries when the latter restricted 
to the space of logarithmic scale factors. At the same time, $SO(1,9)$ is also very 
closely related to the indefinite (Cartan-Killing) metric on the Cartan subalgebra of 
the hyperbolic Lie algebra $E_{10}$~\cite{Damour:2002et}. 

The emergence of the DeWitt metric and of its Lorentzian invariance group 
as the principal symmetry in the present context is rather remarkable: 
it means that our new representations are `higher spin' {\em not} in physical space-time, 
but rather in some variant (or generalization) of Wheeler-DeWitt superspace! 
As a consequence, the connection with physical spin (in $D=11$ dimensions) 
is slightly obscured: although our representations are larger than the usual 
spin-$\frac12$ and spin-$\frac32$ representations, they are not necessarily higher 
spin from the point of view of the usual spatial rotation group $SO(10)$ in $D=11$
dimensions. Such an interpretation of our results is completely in line with the 
expected de-emergence of space (and time) near the singularity: in the Planck regime, 
where classical space-time ceases to exist, conventional notions of spin may also 
become meaningless. In such a `pre-geometric phase'  only the duality symmetry
should remain as a defining feature of the theory, along the lines of the M Theory 
proposal of \cite{Damour:2002cu}.

Even though we restrict to $K(E_{10})$ in this paper for concreteness, we expect 
our methods and results to apply more generally. In fact, it is easy to see that the same 
formulas hold for the whole $E_{11-D}$ series, including affine $E_9$ and the 
Lorentzian Kac--Moody algebra $E_{11}$. 

This article is structured as follows. We first review the necessary algebraic framework for describing $E_{10}$ and $K(E_{10})$, consistency conditions for $K(E_{10})$ representations as well as the known elementary finite-dimensional unfaiithful representations of $K(E_{10})$ in section~\ref{sec:prelim}. Then we give the implementation of these representation as quantum operators in section~\ref{sec:qops}. A convenient parametrisation of the $E_{10}$ root lattice in section~\ref{sec:Par} allows us to derive new forms of the consistency conditions for $K(E_{10})$ representations. In section~\ref{sec:HS}, we construct our new higher-spin 
representations and study their properties; their inequivalence with the known
representations is established in section~\ref{sec:Truncations} by considering their
decompositions under $SO(16)$. In the final section~\ref{sec:WG}, we investigate 
the realization of the Weyl group.

\section{Algebraic preliminaries}
\label{sec:prelim}

In this section, we review some basic definitions and properties of the maximally
extended hyperbolic Kac-Moody algebra $E_{10}$ and its maximal compact 
subalgebra $K(E_{10})$, as well as the known unfaithful spinorial representations
(associated to `spin-$\frac12$' and `spin-$\frac32$', respectively). Referring to 
\cite{Damour:2006xu} for further details, we first give all relevant expressions 
in $\mf{so}(10)$ covariant form.

\subsection{$E_{10}$ and $K(E_{10})$}

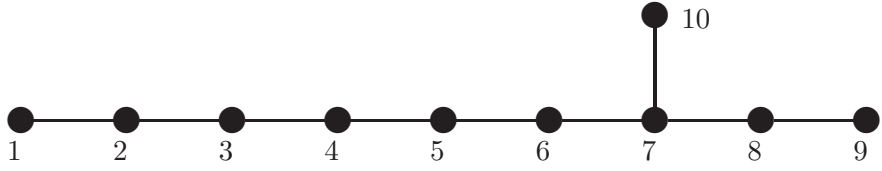
\begin{figure}
\begin{center}
\scalebox{1}{
\begin{picture}(340,60)
\put(5,-5){$1$} \put(45,-5){$2$} \put(85,-5){$3$}
\put(125,-5){$4$} \put(165,-5){$5$} \put(205,-5){$6$}
\put(245,-5){$7$} \put(285,-5){$8$} \put(325,-5){$9$}
\put(260,45){$10$} \thicklines
\multiput(10,10)(40,0){9}{\circle*{10}}
\multiput(15,10)(40,0){8}{\line(1,0){30}}
\put(250,50){\circle*{10}} \put(250,15){\line(0,1){30}}
\end{picture}}
\caption{\label{fig:E10}\sl Dynkin diagram of $E_{10}$ with numbering of nodes.}
\end{center}
\end{figure}

The hyperbolic Kac--Moody algebra $E_{10}$ is defined in terms of generators and relations that are encoded in its Dynkin diagram displayed in figure~\ref{fig:E10}. We consider $E_{10}$ in its split real form which is an infinite-dimensional simple Lie algebra. Associating to each node of the Dynkin diagram a standard Chevalley triple $(e_i,h_i,f_i)$ ($i=1,\ldots,10$) the algebra is generated by all multiple commutators of the elements of the elementary triples,
 subject to the relations (see e.g.~\cite{Kac})
\begin{subequations}
\label{e10relations}
\begin{align}
\lb e_i,f_j \rb &= \delta_{ij} h_i, &\lb h_i, e_j\rb &= A_{ij} e_j,& \lb h_,f_j\rb &=-A_{ij}f_j,&\\
\label{e10serre}
(\ad{e_i})^{1-A_{ij}} e_j &=0,& (\ad{f_i})^{1-A_{ij}} f_j &=0.&
\end{align}
\end{subequations}
Here, $A_{ij}$ is the Cartan matrix of $E_{10}$ with values $A_{ii}=2$ on the diagonal and the off-diagonal entries $A_{ij}=-1$ (for nodes joined by a direct line) or $A_{ij}=0$ (for nodes between which there is no line). The generators $e_i$ and $f_i$ are referred to as positive and negative step operators, respectively. The relations (\ref{e10relations}) imply in particular that each triple $(e_i,h_i, f_i)$ by itself forms a subalgebra isomorphic to $\mf{sl}(2,\reals)$; the relations (\ref{e10serre}) are called the Serre relations and determine the ideal that has to be quotiented out from the free Lie algebras generated by the $e_i$ and $f_i$.  The Cartan subalgebra $\mf{h}$ is the linear span of the generators $h_i$ and of real dimension $10$. The adjoint action of an element $h\in\mf{h}$ on $E_{10}$ can be diagonalised, and as usual the eigenvalues 
are the roots $\alpha\in\mf{h}^*$ of the algebra with eigenspaces
\begin{align}
(E_{10})_\alpha = \left\{ x\in E_{10} \,\middle|\, \lb h,x\rb = \alpha(h) x \quad\text{for all $h\in\mf{h}$}\right\}.
\end{align}
The eigenspaces are of finite (but exponentially growing) dimension 
${\rm mult}(\alpha) = \dim (E_{10})_\alpha$. The lattice of roots $Q(E_{10})\cong \oplus_{i=1}^{10} \ints \alpha_i$ provides a grading of $E_{10}$:
\begin{align}
\label{e10grading}
\big[ (E_{10})_\alpha, (E_{10})_\beta\big] \subset (E_{10})_{\alpha+\beta}.
\end{align}

On the Lie algebra $E_{10}$ one can define a standard involution $\omega$, which 
is called the Chevalley involution and acts by
\begin{align}
\omega(e_i) = -f_i,\quad\omega(h_i)=-h_i.
\end{align}
It is real linear and satisfies for all $x,y\in E_{10}$ the invariance property
\begin{align}
\omega(\omega(x))=x,\quad  \omega\left(\lb x,y\rb\right) = \lb \omega(x),\omega(y)\rb.
\end{align}
The Lie algebra $E_{10}$ can then be divided into eigenspaces of $\omega$
\begin{align}
E_{10} = K(E_{10}) \oplus   K(E_{10})^\perp  %\mf{p}
\end{align}
with
\begin{subequations}
\begin{align}
K(E_{10}) &=\left\{ x\in E_{10}\,\middle|\, \omega(x) =x\right\},\\
%\mf{p} 
K(E_{10})^\perp &=\left\{ x\in E_{10}\,\middle|\, \omega(x) =-x\right\}.
\end{align}
\end{subequations}
The space $K(E_{10})$ of $\omega$-invariant elements is called the {\em compact 
subalgebra} of $E_{10}$ and plays the lead role in this paper. It is called `compact' 
because its Killing metric, inherited from the indefinite bilinear form of $E_{10}$, is negative definite. In particular, the Cartan subalgebra $\mf{h}$ is not part of $K(E_{10})$.

There is also a presentation of $K(E_{10})$ in terms of generators and relations 
that follows directly from (\ref{e10relations}). Because
\begin{align}
\label{kgens}
x_i =e_i-f_i
\end{align}
is the only invariant combination in each standard triple (corresponding to the 
compact $\mf{so}(2)\subset\mf{sl}(2,\reals)$), one can show that $K(E_{10})$ is the 
Lie algebra generated by the ten $x_i$, subject to the relations~\cite{Berman}
\begin{align}
\label{ke10relations}
\sum_{k=0}^2 C_{ij}^{(k)} (\ad{x_i})^k x_j =0,
\end{align}
where
\begin{subequations}
\begin{align}
C_{ij}^{(0)} &= 0,&C_{ij}^{(1)} &= 1,&C_{ij}^{(2)} &= 0, &&\text{if $A_{ij}=0$},&\\
C_{ij}^{(0)} &= 1,&C_{ij}^{(1)} &= 0,&C_{ij}^{(2)} &=1, &&\text{if $A_{ij}=-1$}.&
\end{align}
\end{subequations}
The first line reflects the fact that the $\mf{so}(2)$ algebras of nodes that are not connected in the Dynkin diagram commute. The second line gives the non-trivial relations between nodes connected by a single line, corresponding to $\mf{so}(3)$.\footnote{When there are multiple lines in the Dynkin diagram, the relations (\ref{ke10relations}) become higher order. For example, for $AE_3$, there is a relation between the nodes $2$ and $3$ connected by a quadruple line
\begin{equation*}
\lb x_2, \lb x_2,\lb x_2,x_3\rb\rb\rb + 4 \lb x_2,x_3\rb=0.
\end{equation*}} Observe that, as for all algebras with at most one line between any pair of nodes, 
at most double commutators appear in these relations for $K(E_{10})$. The above relations will play a key role in the remainder. More specifically, we will construct concrete sets of 
fermionic bilinears made from `higher spin' fermions which satisfy (\ref{ke10relations}). 
Berman's Theorem~\cite{Berman} then guarantees that each such set provides a realization 
of the algebra $K(E_{10})$.

The Lie algebra $K(E_{10})$ is not a Kac--Moody algebra~\cite{Kleinschmidt:2005bq}. 
In particular, there is no grading of $K(E_{10})$ by finite-dimensional root spaces 
analogous to (\ref{e10grading}). Instead (\ref{e10grading}) induces a {\em filtered 
structure} on $K(E_{10})$~\cite{Damour:2006xu}. Because the generators of $K(E_{10})$ 
are always combinations of positive and negative step operators of $E_{10}$ we 
denote the corresponding subspaces of $E_{10}$ by
\begin{align}
\label{ke10rts}
(K(E_{10}))_\alpha \equiv  (K(E_{10}))_{-\alpha} 
\equiv (E_{10})_\alpha \oplus (E_{10})_{-\alpha}
\end{align}
Then it is immediately evident that
\begin{align}
\label{Filter}
\big[ (K(E_{10}))_\alpha,(K(E_{10}))_\beta\big] \subset (K(E_{10}))_{\alpha+\beta} \oplus (K(E_{10}))_{\alpha-\beta}.
\end{align}
We will often speak loosely of a root $\alpha$ for $K(E_{10})$ when we mean the space (\ref{ke10rts}) and associated $K(E_{10})$ generators $J(\alpha)$, keeping in mind that commutators of such generators can go `up and down' according to (\ref{Filter}).

\subsection{Unfaithful spinors in $\mf{so}(10)$ covariant form}

A useful way of presenting the algebras $E_{10}$ and $K(E_{10})$ is in terms of a so-called level decomposition~\cite{Damour:2002cu}. This means that one chooses a subalgebra of $E_{10}$ obtained by deleting nodes from the Dynkin diagram and then decomposes the adjoint representation of $E_{10}$ into representations of that subalgebra. We restrict here to a decomposition of $E_{10}$ under its $\mf{sl}(10,\reals)$ subalgebra obtained by deleting node number $10$ from figure~\ref{fig:E10}. Then all $E_{10}$ generators can be represented in an $\mf{sl}(10,\reals)$ covariant form. At the lowest non-negative levels, this leads to the following generators~\cite{Damour:2002cu}:
\begin{align}
K^a{}_b,\quad E^{a_1a_2a_3}=E^{[a_1a_2a_3]},\quad E^{a_1\ldots a_6}= E^{[a_1\ldots a_6]},\,\ldots
\end{align}
where $K^a{}_b$ denotes the adjoint of $\mf{gl}(10,\reals)=\mf{sl}(10,\reals)\oplus \reals$ and all tensor indices take values $a,b=1,\ldots,10$. The commutation relations (\ref{e10relations}) could then be restated in an $\mf{sl}(10,\reals)$ covariant form involving only the first two generators $K^a{}_b$ and $E^{a_1a_2a_3}$.

Our main interest here is the corresponding structure for $K(E_{10})$. The image of 
$\mf{sl}(10,\reals)$ in $K(E_{10})$ is $\mf{so}(10)$ and the images of the generators above are
\begin{align}
\label{ke10lev1}
J^{ab} = K^a{}_b - K^b{}_a,\quad J^{a_1a_2a_3} = E^{a_1a_2a_3}+\omega(E^{a_1a_2a_3}),\,\ldots
\end{align}
where the $J^{ab}$ generate the compact subalgebra $\mf{so}(10)\subset \mf{sl}(10,\reals)$ 
and $J^{a_1a_2a_3}$ is the first generator in the infinite-dimensional extension 
$K(E_{10})$ of $\mf{so}(10)$. Importantly, all defining relations (\ref{ke10relations}) 
can be restated in an $\mf{so}(10)$ covariant form that involves only the generators 
of (\ref{ke10lev1}). Consequently, we can decompose the whole 
$K(E_{10})$ algebra in terms of $\mf{so}(10)$ tensors.

Representations of $K(E_{10})$ can then be constructed by combining $\mf{so}(10)$ representations with an action of $J^{a_1a_2a_3}$ in such a way that the
$\mf{so}(10)$ version of the relation (\ref{ke10relations}) is satisfied.
In the papers~\cite{Damour:2005zs,deBuyl:2005mt,Damour:2006xu} this was 
done for the two examples of primary physical interest, namely the `spin-$\frac32$' 
vector spinor (gravitino) and the `spin-$\frac12$' Dirac spinor (supersymmetry parameter) 
of $D=11$ supergravity (in our previous work, we alternatively referred to the 
vector spinor representation as the `Rarita-Schwinger' representation).

The Dirac-representation of $K(E_{10})$ is made up of a $32$ component Majorana (real) spinor 
$\chi$ of $\mf{so}(10)$. The action of the generators (\ref{ke10lev1}) on $\chi$ is given by
\begin{align}
\label{DSlin}
J^{ab} \chi = \frac12 \Gamma^{ab} \chi,\quad 
J^{a_1a_2a_3} \chi = \frac12 \Gamma^{a_1a_2a_3} \chi.
\end{align}
Here, we follow the conventions of~\cite{Damour:2006xu} where the $\mf{so}(10)$
Clifford algebra is generated by the $32\times 32$ $\Gamma$-matrices $\Gamma^a$ 
which are symmetric and real, and we make use of the standard notation
$\Gamma^{a_1\ldots a_p} = \Gamma^{[a_1}\cdots \Gamma^{a_p]}$.~\footnote{The 
spinor indices $\alpha,\beta,\dots = 1,...,32$ will usually not be written out explicitly.}
As one can easily see, the representation of $K(E_{10})$ on the Dirac spinor
generated by multiple commutation of the above generators (\ref{DSlin}) is simply
spanned by the anti-symmetric combinations of $\Gamma$-matrices, which combine
to a basis of the compact Lie algebra $\mf{so}(32)$.

By contrast, the $K(E_{10})$ vector spinor $\psi_a$ has $320$ real components 
and consists of \emp{two} irreducible $\mf{so}(10)$ representations: In 
$\mf{so}(10)$ the $\Gamma$-trace $\Gamma^a \psi_a$  is an irreducible 
component by itself and $\psi_a$ decomposes into a $\Gamma$-traceless part and a pure trace. This is no longer true in $K(E_{10})$ and there the full $\psi_a$ including the $\Gamma$-trace is an irreducible representation. The action of the generators (\ref{ke10lev1}) is given by~\cite{Damour:2005zs,deBuyl:2005mt}
\begin{subequations}
\label{vs01}
\begin{align}
J^{ab} \psi_c &= \frac12 \Gamma^{ab}\psi_c + 2\delta_c^{[a}\psi^{b]},&\\
J^{a_1a_2a_3} \psi_c &= \frac12 \Gamma^{a_1a_2a_3}\psi_c + 4\delta_c^{[a_1}\Gamma^{a_2}\psi^{a_3]} - \Gamma_{c}{}^{[a_1a_2}\psi^{a_3]}.&
\end{align}
\end{subequations}
(Because indices here are raised and lowered with the $\mf{so}(10)$ invariant 
$\delta^{ab}$, their position does not really matter.) By self-consistency of the 
representation one can deduce the action of the higher level generators from 
these rules. The resulting formulas for levels three and  four can be found 
in~\cite{Damour:2005zs,deBuyl:2005mt,Henneaux:2008nr}. 

Both the Dirac spinor and the vector spinor are {\em finite-dimensional} representations 
of the infinite-dimensional algebra $K(E_{10})$, and therefore \emp{unfaithful}. Concretely,
this means that there exists an infinite number of combinations of $K(E_{10})$ generators 
that are represented trivially in this the representation.  As a consequence, the algebra 
$K(E_{10})$ is {\em not} simple: it has non-trivial ideals generated by those
combinations that are represented trivially on the representation. 
An important notion for unfaithful representations is that of the \emp{quotient algebra} that is obtained by quotienting $K(E_{10})$ by the non-trivial ideal. We will be interested in constructing `more faithful' representations that are in particular not obtained by taking tensor products of known ones.

For both representations one can define invariant bilinear forms on the respective
representation spaces~\cite{Damour:2006xu}. For the Dirac spinor one has
\begin{align}\label{bilD}
\left(\chi\middle|\chi\right)_{\text{DS}} = \chi_\alpha \chi_\alpha \equiv \chi^T \chi
\end{align}
and on the vector spinor
\begin{align}
\label{bilvs}
\left(\psi\middle|\psi\right)_{\text{VS}} = - \, \psi_a^T\Gamma^{ab} \psi_b.
\end{align}
Whereas the bilinear form (\ref{bilD}) is manifestly positive, the one for the
vector spinor is indefinite on the representation space;
this is possible because $\psi_a$ is a combination of {\em two} irreducible representations
of $\mf{so}(10)$. In principle, one can treat the (classical) spinor fields as either commuting (symmetric) or anti-commuting $c$-numbers (anti-symmetric $(\cdot|\cdot)$). However, for
the above bilinear forms not to vanish we must assume that the classical fermion 
variables are treated as commuting $c$-numbers.~\footnote{By contrast, but for the very
 same reason, one must take the fermions as {\em anti}-commuting in the fermionic
 Lagrangian $\propto \psi_a^T\Gamma^{ab} \partial_t\psi_b$ \cite{Damour:2006xu}.}

The invariance of the bilinear forms means that they are compatible with the 
action of $K(E_{10})$. For the Dirac representation this follows trivially from 
the anti-symmetry of the representation matrices; for the vector spinor
one verifies by explicit computation~\cite{Damour:2006xu} that 
\begin{align}
\left(x\cdot\psi\middle|\psi\right)_{\text{VS}} +\left(\psi\middle|x\cdot\psi\right)_{\text{VS}} =0\quad
\text{for any $x\in K(E_{10})$.}
\end{align}
As also shown there the existence of such a form is special to the hyperbolic 
extension (so, for instance, there is no corresponding bilinear form on $K(E_{11})$).

\section{$K(E_{10})$ via quantum operators}
\label{sec:qops}

\subsection{Quantum brackets}

The classical supergravity action induces a canonical Dirac bracket on the 
vector spinor~\cite{Damour:2006xu}. We will pass to the quantum theory right away,
by replacing the classical fields $\chi$ and $\psi_a$ by quantum operators
$\qchi$ and $\qpsi_a$ (as well as  setting $\hbar =1$ as usual), the classical 
bracket gives rise to the canonical anti-commutation relations 
in the manifestly $\mf{so}(10)$ covariant form 
\begin{align}\label{CB1}
\big\{ \qpsi_{a\alpha}, \qpsi_{b\beta} \big\} 
= \delta_{ab}\delta_{\alpha\beta} -\frac19 (\Gamma_a\Gamma_b)_{\alpha\beta},
\end{align}
for the vector spinor, where we have written out the $32$-component spinor indices
as well and adopted a convenient normalization of the fermion operators. 
Similarly, the bracket between two Dirac spinors becomes\footnote{We note
 that there is no such (propagating) Dirac spinor in $D=11$ supergravity, but this is the 
 standard result that one would obtain for a spin-$\frac12$(Majorana) field in any dimension. } 
\begin{equation}\label{CB2}
\ld \qchi_\alpha , \qchi_\beta \rd = \delta_{\alpha\beta}\;\; ,
\end{equation}
again with a convenient normalization. (The standard normalization for a Majorana spinor would include a factor of $\frac12$ in the canonical Dirac bracket that we have absorbed here into the definition of $\chi$.)
 
It is straightforward to find the quantum realizations of the generators $J^{ab}$
and $J^{abc}$ in terms of the quantum operators $\qchi$ and $\qpsi_a$; we have 
\begin{equation}
\hJ^{ab} = \frac14 \qchi^T \Gamma^{ab} \qchi\; , \quad
\hJ^{abc} = \frac14 \qchi^T \Gamma^{abc} \qchi
\end{equation}
for the Dirac spinor, and
\begin{eqnarray}\label{JSO10}
\hJ^{ab} &=& - \,\qpsi^a\qpsi^b - \frac14 \,\qpsi^c \Gamma^{ab} \qpsi^c 
        + \, \frac14 \qpsi^c \Gamma^c \Gamma^{ab} \Gamma^d \qpsi^d  \nn\\
\hJ^{abc} &=& - \, 3 \qpsi^{[a} \Gamma^b \qpsi^{c]} -\, \frac14 \qpsi^e \Gamma^{abc} \qpsi^e
           -\,  \frac14 \qpsi^e \Gamma^e \Gamma^{abc} \Gamma^f \qpsi^f \\
 \hJ^{a_1\cdots a_6} &=& 15 \qpsi^{[a_1} \Gamma^{a_2\cdots a_5} \qpsi^{a_6]}
 - \,\frac14 \qpsi^b \Gamma^{a_1\cdots a_6} \qpsi^b
 + \, \frac14 \qpsi^b\Gamma^b \Gamma^{a_1\cdots a_6} \Gamma^c \qpsi^c          \nn
\end{eqnarray}
for the vector spinor. The `unusual'  trace terms on the r.h.s. are due to the extra
term on the r.h.s. of the bracket (\ref{CB1}) above (without this term, the first two terms 
on the r.h.s., respectively, would generate the orbital and the spin parts of the rotation). 
Readers are invited to check that these generators do satisfy the defining 
relations (\ref{ke10relations}).

\subsection{Breaking $\mf{so}(10)$ covariance}

An important insight of \cite{Damour:2009zc}  was that the analysis of the
fermionic billiards becomes simpler if one writes all expressions in terms 
of a redefined vector spinor. This different form of the $K(E_{10})$ 
vector spinor $\psi_a$ is obtained by breaking the manifest $\mf{so}(10)$ invariance. 
For this purpose {\em we suspend the  summation convention for $SO(10)$
vector indices~\footnote{But not for $\mf{so}(10)$ spinor indices.} from now on} and define
\begin{align}
\label{newvs}
\phi^\ta_\alpha \equiv \Gamma^a_{\alpha\beta}\psi^a_{\beta}\quad\text{\bf (no sum on $a$!)}
\end{align}
Here, we have switched to a different font in order to indicate that the vector indices $\ta,\tb,\dots$
are no longer associated to $\mf{so}(10)$ but rather to $\mf{so}(1,9)$, as will be explained
below. This form of arranging the $320$ vector spinor components is better suited for analysing the effect of cosmological billiard reflections in a first~\cite{Damour:2009zc} or second-quantized~\cite{Kleinschmidt:2009cv,Kleinschmidt:2009hv,Damour:2011yk} form. 

The $K(E_{10})$ invariant bilinear form (\ref{bilvs}) in the $\phi^\ta$ coordinate reads
\begin{align}
\label{bilvsnew}
\left(\phi\middle|\phi\right)_{\text{VS}}&=
-\, \sum_{a,b} \phi^{a\,T}\Gamma_a\Gamma^{ab}\Gamma_b \phi^b,&\nn\\
&= -\, \sum_{\ta\neq \tb}  \phi^{\ta\,T} \phi^\tb = \sum_{\ta,\tb} \phi^{\ta\,T} G_{\ta\tb} \phi^\tb,
\end{align}
Here, we have introduced the metric $G_{\ta\tb}$ with
\begin{equation}
\sum_{\ta,\tb} G_{\ta\tb} v^\ta w^\tb \equiv \sum_\ta v^\ta w^\ta - 
        \left(\sum_\ta v^\ta\right)\left(\sum_\tb w^\tb\right)
\end{equation}
and written out the sums explicitly. The inverse metric is
\begin{equation}\label{inverseDW}
\sum_{\ta,\tb} G^{\ta\tb} v_\ta w_\tb \equiv \sum_\ta v^\ta w^\ta - 
        \frac19\left(\sum_\ta v_\ta\right)\left(\sum_\tb w_\tb\right)
\end{equation}
Importantly, {\em this metric is Lorentzian}; in fact, it is just the DeWitt metric 
that appears in the bosonic dynamics of the scale factors in a cosmological billiard approximation~\cite{Damour:2000wm,Damour:2002et} (which itself is the restriction of the
full DeWitt metric on the moduli space of 3-geometries in canonical gravity).
Thus, even though the redefinition (\ref{newvs}) breaks the $\mf{so}(10)$ covariance 
on the vector spinor $\psi^a$, it introduces a new $\mf{so}(9,1)$ structure on the 
redefined vector spinor $\phi^\ta$. 
At the same time, we have 
replaced the positive definite metric $\delta_{ab}$ by the indefinite metric
$G_{\ta\tb}$. Below we will re-instate the summation convention with respect
to these indices, such that for instance 
\begin{equation}
v_\ta w^\ta \equiv  G^{\ta\tb} v_\ta w_\tb \equiv\sum_{\ta,\tb} G^{\ta\tb} v_\ta w_\tb 
\end{equation}
and so on.

In terms of the redefined fermions (\ref{newvs}) the bracket (\ref{CB1}) becomes
\begin{align}
\label{qbracket}
%\frac{i}2 
\ld \qphi_{\alpha}^\ta, \qphi_{\beta}^\tb \rd = G^{\ta\tb}\delta_{\alpha\beta},
\end{align}
Note that it is the {\em inverse} metric $G^{\ta\tb}$ (\ref{inverseDW}) which appears on 
the r.h.s., and this is also the reason why the vector index on $\phi^\ta$ is naturally placed upstairs. The  Fock space realization of (\ref{qbracket}) requires replacing the $320$ real 
operator components $\phi^\ta_\alpha$ by $160$ complex linear combinations 
in the standard way, that is, $160$ creation and $160$ annihilation operators 
\cite{Kleinschmidt:2009cv,Kleinschmidt:2009hv}; the resulting fermionic Fock
space has dimension $2^{160}$.
Remarkably, although constructed from a set of $SO(10)$ covariant fermion operators,
this Fock space is an {\em indefinite metric} state space. As explained in 
the introduction, this indefiniteness here is not due to the indefiniteness of the 
metric of the original 11-dimensional physical space-time, but linked to the 
indefiniteness of the DeWitt metric, which is a consequence of the lower 
unboundedness of the Einstein-Hilbert action under variations of the conformal factor.

\subsection{Action of $K(E_{10})$ on $\qphi^\ta$}

We now want to describe how the action of $K(E_{10})$ on $\qphi^\ta$ is implemented 
on the state space through the quantum bracket (\ref{qbracket}).  For any fixed 
$x\in K(E_{10})$ we write the action of it on the redefined $\qphi^\ta$ as
\begin{align}
\label{kact}
\delta_x \qphi^\ta.
\end{align}
Abstractly, one can define this action by the operator
\begin{align}
\label{qx}
\hat{x} = \frac12 
G_{\ta\tb}\qphi^\ta \delta_x\qphi^\tb 
=\frac12 \left(\qphi\middle| x\cdot\qphi\right)_{\text{VS}}
\end{align}
for any $x\in K(E_{10})$, where $\delta_x\qphi^\tb$ is the linear action (\ref{kact}) 
of $x$ on $\phi^\tb$, but now realized by the operator $\hat{x}$. The last equality reveals the operator as the matrix obtained from the bilinear form (\ref{bilvsnew}). The operator $\hat{x}$ implements the action on $\qphi^\ta$ through
\begin{align}
\lb \hat{x},\qphi^\tc \rb = 
\lb \frac12 
G_{\ta\tb}\qphi^\ta \delta_x\qphi^\tb, \qphi^\tc \rb = -\delta_x\qphi^\tc.
\end{align}
To check the validity of this assertion we need to ascertain that $\delta_x$ is self-adjoint with respect to the bilinear form; this follows, however, from the invariance of the bilinear form. A formula of the type (\ref{qx}) will always be true when one has a classical $K(E_{10})$ representation $v^{\mathcal{A}}$ (where $\mathcal{A}$ labels the components), this representation possesses an invariant bilinear form with metric $\mathcal{G}_{\mathcal{A}\mathcal{B}}$ and the canonical (anti-)bracket on $v^{\mathcal{A}}$ is $\lb v^{\mathcal{A}}, v^{\mathcal{B}}\rb=\mathcal{G}^{\mathcal{A}\mathcal{B}}$.

\section{Explicit parametrisation in terms of generators}
\label{sec:Par}

Even though the formula (\ref{qx}) realises $K(E_{10})$ via quantum operators on the Fock space, it is rather abstract compared to the explicit low level results (\ref{vs01}). In this section, we introduce an explicit formula for the action of an arbitrary generator of $K(E_{10})$ acting on the Dirac- or vector spinor. This makes recourse to the root space decomposition (\ref{e10grading}) of $E_{10}$.

\subsection{Roots in wall basis and $\Gamma$ matrices}

The roots $\alpha$ of $E_{10}$ can be parametrised explicitly in terms of the 
so-called `wall basis' that also plays a role in cosmological billiards~\cite{Damour:2002et}. Concretely, we write a root $\alpha$ as\footnote{Below we will also denote
  the components of a root $\alpha$ by $\alpha_\ta\equiv p_\ta\,,\, \beta_\ta\equiv q_\ta$, {\it etc.}}
\begin{align}\label{alphap}
\alpha = \sum_{\ta=1}^{10} p_\ta e^\ta =(p_1,p_2,\ldots, p_{10})
\end{align}
where $e^\ta$ are basis vectors of the Cartan subalgebra whose inner product matrix 
equals the inverse of the DeWitt metric of~(\ref{bilvsnew}) 
\begin{align}
e^\ta\cdot e^\tb = G^{\ta\tb}.
\end{align}
The simple roots of $E_{10}$ can then be written explicitly as 
\begin{subequations}
\label{wallbs}
\begin{align}
\alpha_1 &= (1,-1,0,\ldots,0),&\\
\alpha_2 &= (0,1,-1,0,\ldots,0),&\\
\vdots&&\\
\alpha_9 &= (0,\ldots,0,0,1,-1),&\\
\alpha_{10} &= (0,\ldots,0,1,1,1).&
\end{align}
\end{subequations}
Then it is easy to check that $\alpha_i\cdot\alpha_j = A_{ij}$ gives back the $E_{10}$ Cartan 
matrix via
\begin{equation}
\alpha\cdot\beta = \sum_{\ta,\tb} 
G^{\ta\tb} p_\ta q_\tb 
\end{equation}
where $\alpha = \sum p_\ta e^\ta$ and $\beta = \sum q_\ta e^\ta$. To any (not necessarily real) 
root $\alpha$ in (\ref{alphap}) we now associate a $\Gamma$-matrix through
\begin{align}
\label{GAdef}
\Gamma(\alpha) \equiv \Gamma_1^{p_1}\cdots\Gamma_{10}^{p_{10}}.
\end{align}
For example, we 
have $\Gamma(\alpha_1)=\Gamma^{12}$ and $\Gamma(\alpha_{10}) = \Gamma^{8\,9\,10}$. 
This definition only depends on the (integer) exponents $p_\ta\mod 2$.  

The following elementary identities will be crucial
\begin{eqnarray}
\label{gammaids}
\Gamma(\alpha)\Gamma(\beta) &=& (-1)^{\alpha\cdot\beta} \Gamma(\beta)\Gamma(\alpha),
\nn\\
\Gamma(\alpha)^T &=& (-1)^{\frac12\alpha\cdot\alpha} \Gamma(\alpha),\quad
\big( \Gamma(\alpha)\big)^2 = (-1)^{\frac12 \alpha\cdot\alpha}
\end{eqnarray}
Furthermore, for $\alpha=\sum_\ta p_\ta e^\ta$ and $\beta=\sum_\tb q_\tb e^\tb$ 
we define the quantity
\begin{align}
\eps_{\alpha,\beta} = (-1)^{\sum_{\ta<\tb} q_\ta p_\tb}
\end{align}
such that
\begin{equation}
\Gamma(\alpha) \Gamma(\beta) = \eps_{\alpha,\beta} \,\Gamma(\alpha \pm \beta)
\end{equation}
We stress that these relations are only valid because in the exponents 
we can compute mod 2: for any two roots $\alpha$ and $\beta$ associated with 
10-tuples $(p_1,\dots,p_{10})$ and $(q_1,\dots,q_{10})$, respectively,  we have
\begin{equation}
\sum_{\ta,\tb} G^{\ta\tb} p_\ta q_\tb \equiv 
                               \sum_{\ta,\tb} G_{\ta\tb} p_\ta q_\tb \;\; {\rm mod} \, 2
\end{equation}
because $\sum p_\ta \,,\, \sum q_\ta \in 3 \ints$ for all $E_{10}$ roots.

The quantity $\eps_{\alpha,\beta}$ is the two-cocyle that also appears in the 
string vertex operator construction~\cite{Goddard:1986bp}, as well as in the
Lie algebra structure constants when expressed in a Cartan--Weyl basis. 
For any $E_{10}$ root it satisfies the important relations
\begin{eqnarray}
\eps_{\alpha,\beta}\eps_{\beta,\alpha} &=& (-1)^{\alpha\cdot\beta}\;\; , \quad
\eps_{\alpha,\alpha} = (-1)^{\frac12 \alpha\cdot \alpha} \;\; , \nn\\
\eps_{\alpha,\beta} \, \eps_{\beta + \alpha, \gamma} 
&=& \eps_{\alpha,\beta + \gamma} \, \eps_{\beta,\gamma}
\end{eqnarray}

The dependence on the inner product between the two roots in (\ref{gammaids}) means that for given $\alpha$ and $\beta$ either the commutator or the anti-commutator of $\Gamma$-matrices vanishes:
\begin{subequations}
\label{Jcomms}
\begin{align}
\alpha\cdot\beta\in 2\ints &\Longrightarrow \left\{\begin{array}{rcl} \lb\Gamma(\alpha),\Gamma(\beta)\rb &=&0\\[1mm]
 \ld\Gamma(\alpha),\Gamma(\beta)\rd &=&2\eps_{\alpha,\beta} \Gamma(\alpha\pm\beta)\end{array}\right.\\[2mm]
 \alpha\cdot\beta\in 2\ints+1 &\Longrightarrow \left\{\begin{array}{rcl} \lb\Gamma(\alpha),\Gamma(\beta)\rb &=&2\eps_{\alpha,\beta} \Gamma(\alpha\pm\beta)\\[1mm]
 \ld\Gamma(\alpha),\Gamma(\beta)\rd &=&0\end{array}\right.
\end{align}
\end{subequations}
We have indicated that the resulting $\Gamma$ matrix can be interpreted either as 
that of $\alpha+\beta$ or as that of $\alpha-\beta$; modulo $2$ the difference does 
not matter; this fact is actually crucial for recovering the filtered structure (\ref{Filter})
below. We also note that $\Gamma(\alpha)$ can be proportional to identity 
matrix for certain (imaginary) roots $\alpha$ (e.g., twice the null root of $E_9$, or 
any even multiple of an imaginary root).

Finally we define the \emp{Clifford group} to be the finite group generated 
by $\Gamma(\alpha)$. This is simply the set of all $\Gamma$ matrices and
their antisymmetrised products together with their negatives. It is a finite group of order $2^{11}=2048$.

\subsection{Commutation relations}

Labelling the Serre type generators (\ref{kgens}) of $K(E_{10})$ by their roots according to
\begin{align}
x_i \equiv J(\alpha_i)
\end{align}
with the ten simple roots $\alpha_i$ as in (\ref{wallbs}), one can summarise the defining relations (\ref{ke10relations}) as follows. For any
pair of {\em real roots} $\alpha$ and $\beta$ (hence $\alpha^2 = \beta^2 = 2$) 
obeying $\alpha\cdot\beta = \pm 1$ or $=0$ we have
\begin{subequations}
\label{ke10J}
\begin{align}
[J(\alpha)\, , J(\beta) ] &= \eps_{\alpha,\beta} J(\alpha\pm\beta) 
         &&  \mbox{ if $\alpha\cdot\beta = \mp 1$} \\[2mm]
[J(\alpha)\, , J(\beta) ] &= 0 
         && \mbox{ if $\alpha\cdot\beta = 0 $} 
\end{align}
\end{subequations}
These relations are in particular valid for the simple roots $\alpha_i$ of $E_{10}$,
and it is thus straightforward to check all required relations in (\ref{ke10relations})
are satisfied. In particular, for any adjacent roots $\alpha$ and $\beta$
on the Dynkin diagram we have
\begin{equation}
\big[ J(\alpha) , \big[ J(\alpha), J(\beta) \big]\big]
= \eps_{\alpha,\alpha+\beta} \eps_{\alpha,\beta} J(\beta) = - J(\beta)
\end{equation}
Starting from any set of real roots, one can in this way cover a larger and larger set 
of real roots step by step, and in this way arrive at a concrete realization of an infinite
set of $K(E_{10})$ elements.

\subsection{Bilinear quantum operators for $K(E_{10})$}

We now set up a formalism to handle representations of $K(E_{10})$ for higher-spin fermionic operators  $\qphi^\cA$ where $\cA \equiv \{\ta_1\cdots \ta_k\}$ is a $k$-tuple of $SO(1,9)$ tensor indices in some 
(not necessarily irreducible) representation. We will restrict to symmetric $k$-tuples 
$(\ta_1\ldots \ta_k)$ corresponding to what we refer to as `Spin-($\frac{k+2}2)$' 
but the formalism applies more generally. For the quantum operators $\qphi^\cA$ 
we assume canonical commutation relations of the form
\begin{align}\label{qB1}
\ld \qphi^\cA_\alpha, \qphi^\cB_\beta\rd = G^{\cA\cB}\delta_{\alpha\beta},
\end{align}
where $G^{\cA\cB}$ denotes some non-degenerate and $SO(1,9)$-invariant metric. The class of operators therefore includes the Dirac spinor $\qchi$ (with $\cA$ absent, and $X=1$) 
and the vector spinor $\qphi^\ta$ with $G^{\cA\cB}= G^{\ta\tb}$ (cf. (\ref{qbracket})). It is important to note that for any $SO(1,9)$ invariant metric $G^{\cA\cB}$ one obtains a $K(E_{10})$ invariant metric on the space of $\qphi^\cA_\alpha$ whenever $\qphi^\cA_\alpha$ can be turned into a representation of $K(E_{10})$, the criterion for which we will now discuss.

{}From (\ref{qx}) we see that we require bilinear operators in $\qphi^\cA$ in order to write the $K(E_{10})$ generators. For this reason we 
consider the following expressions
\begin{align}
\label{VSbils}
\hA = \sum_{\cA,\cB} X_{\cA\cB} S^{\alpha\beta} \qphi^\cA_\alpha \qphi^\cB_\beta\quad ,\qquad
\hB = \sum_{\cC,\cD} Y_{\cC\cD} T^{\gamma\delta} \qphi^\cC_\gamma \qphi^\cD_\delta.
\end{align}
If we exclude the identity operator, the properties 
of (\ref{qB1}) imply that the product matrices $X\otimes S$ and $Y\otimes T$ must
be antisymmetric under simultaneous interchange of the index pairs $(\cA,\alpha)$
and $(\cB,\beta)$~\footnote{Otherwise, the resulting expression would reduce to
  the unit operator or simply vanish.}, hence we must have
 $X_{\cA\cB}S^{\alpha\beta} = - X_{\cB\cA} S^{\beta\alpha}$, 
and similarly for $\hB$. This implies that either 
$X_{\cA\cB}=-X_{\cB\cA}$, $S^{\alpha\beta}=+S^{\beta\alpha}$ or 
$X_{\cA\cB}=+X_{\cB\cA}$, $S^{\alpha\beta}=-S^{\beta\alpha}$. The factorised 
form of the ansatz (\ref{VSbils}) is justified by the linearity of the Lie bracket.

Calculating the commutator of the two bilinears $\hA$ and $\hB$ one finds
\begin{align}
\label{ABcomm}
\lb \hA, \hB\rb = \sum_{\cA,\cB} \qphi^\cA_\alpha \left( \lb X,Y\rb_{\cA\cB} \ld S,T\rd_{\alpha\beta} 
+ \ld X,Y \rd_{\cA\cB} \lb S,T\rb_{\alpha\beta}\right)\qphi^\cB_\beta,
\end{align}
so that the bilinear form of the ansatz is preserved.\footnote{Here, $\lb X,Y\rb_{\cA\cB} = \sum_{\cC,\cD}\lp
X_{\cA\cC}G^{\cC\cD}Y_{\cD\cB} -  Y_{\cA\cC}G^{\cC\cD}X_{\cD\cB}\rp$, etc.} This will
be our `master identity' for the calculations below. Note that this identity also applies 
for Dirac spinors with $X(\alpha)=1/4$ and $S(\alpha)=\Gamma(\alpha)$ for real roots $\alpha$, and then only the second term in parentheses
contributes, explaining again why only {\em anti}-symmetric $\Gamma$-matrices
appear in the Dirac representation as we will now see in more detail.

\subsection{Dirac spinor}

For the Dirac spinor with the anti-commutation relation (\ref{CB2}) the implementation of 
the action of $K(E_{10})$ then follows from the discussion, with the result
\begin{align}
\label{hJDS}
\hJ(\alpha) = \frac14\qchi^T\Gamma(\alpha)\qchi.
\end{align}
The generators of $K(E_{10})$ here are thus parametrised by the positive roots of $E_{10}$,
but there is only a finite number of such operators because we the components of any
root $\alpha$ are only considered modulo 2. In particular, we can verify once again 
the defining relations (\ref{ke10relations}):  writing the above expression not 
in terms of quantum operators but in first quantized form analogous to (\ref{DSlin}) one has 
\begin{align}\label{dchi}
 [J(\alpha) \,,\, \chi] = 
-\frac14\left(\Gamma(\alpha)- \Gamma(\alpha)^T\right)\chi
=-\frac14\left(1-(-1)^{\frac12\alpha\cdot\alpha}\right)\Gamma(\alpha)\chi=-\delta \chi  ,
\end{align}
where we made use of (\ref{gammaids}). Therefore, only anti-symmetric 
$\Gamma$-matrices contribute. The commutation relations of the operators 
$J(\alpha)$ are then determined by (\ref{Jcomms}).

Because all the defining relations hold, one can now generate the full algebra $K(E_{10})$
by successive commutation, but it is immediately obvious that there is an infinite-fold
degeneracy in the sense that the infinitely many elements of $K(E_{10})$ are mapped
to a finite set of operators. In particular, by virtue of (\ref{dchi})
all elements of $K(E_{10})$ associated
to (imaginary) roots with $\alpha^2\in 4\ints$ are represented trivially; moreover, all
elements associated to an imaginary root $\alpha$ act in the same way, even if the 
multiplicity of the root is $> 1$. The lack of faithfulness is thus directly related to the 
fact that the Clifford relation $\Gamma_a^2 = 1$ implies there are only finitely many
different $\Gamma$-matrices, and the Clifford group is a {\em finite} group.

We note at this point that the actual real representation of the Clifford algebra chosen for (\ref{hJDS}) does not really matter as long as it is faithful and represents (\ref{Jcomms}). In this way one can (artificially) increase the dimension of the representation but from the point of view of analysing $K(E_{10})$ this does not really yield interesting new information since the active quotient algebra does not change. Similarly, we do not consider here the possibility of tensoring known representation for very much the same reason. Certainly, taking tensor powers of a single given representation clearly does not change the ideal associated with this representation and hence leaves the quotient unchanged. Taking tensor products between different representations can in principle yield more faithful representations and we leave the investigation of this question to future work.

\subsection{Vector spinor}

To derive the operator representations $\hA$ for the vector spinor and to 
arrive at a simple form of the generators for real roots $\alpha$, we re-write
the generators (\ref{JSO10}) in terms of the new spinor operators (\ref{newvs})
and find
\begin{subequations}
\begin{eqnarray}
\hJ^{ab} &=& - \frac12 \big( \qphi^a - \qphi^b\big) \Gamma^{ab} \big(\qphi^a - \qphi^b\big)
            + \frac14 \sum_{c,d} G_{cd} \qphi^c \Gamma^{ab} \qphi^d ,\\
\hJ^{abc} &=& -      \frac12 \big( \qphi^a + \qphi^b+ \qphi^c\big) 
    \Gamma^{abc} \big(\qphi^a + \qphi^b + \qphi^c\big)   
      + \frac14 \sum_{d,e} G_{de} \qphi^d \Gamma^{abc} \qphi^e    \nn\\    
\end{eqnarray}
\end{subequations}
(remember that the summation convention on $SO(10)$ indices remains suspended). 
Next, using the explicit expressions for the simple roots $\alpha$ of $E_{10}$ in (\ref{wallbs})
it is easy  to see that both formulas can be neatly and compactly re-written in the form 
(\ref{VSbils}) with
\begin{equation}
\label{X32}
X(\alpha)_{\ta\tb} = - \frac12 \alpha_\ta \alpha_\tb + \frac14 G_{\ta\tb}
\end{equation}
As special cases, we obtain the $K(E_{10})$ generators associated to the simple roots 
$\alpha_\ta=(\alpha_i)_\ta$ in the universal form
\begin{align}
\label{Jreal}
\hJ(\alpha_i) =
X_{\ta\tb}(\alpha_i) \, \qphi^\ta\Gamma(\alpha_i)\qphi^\tb.
\end{align}
where we have now re-instated the summation convention for the $SO(1,9)$ indices;
observe that in this expression, any reference to $SO(10)$ has disappeared!
The $SO(1,9)$ tensor $X(\alpha)_{\ta\tb}$ satisfies the two crucial identities
\begin{align}\label{Master}
\big\{ X(\alpha) , X(\beta)  \big\}_{\ta\tb}  &= \frac12 X_{\ta\tb}(\alpha \pm \beta) 
&&    \textrm{if $\alpha\cdot\beta = \mp 1$}  \nn \\[2mm]
\big[ X(\alpha) , X(\beta)   \big]_{\ta\tb}  &= 0   &&  \textrm{if $\alpha\cdot\beta = 0 $}   
 \end{align}
where of course, again $\alpha^2 = \beta^2 =2$. With the identities 
(\ref{Jcomms}) and (\ref{ABcomm}) it is straight-forward to verify the 
equivalent form of the defining $K(E_{10})$ generating relations (\ref{ke10J}).

Remarkably the simple formula  (\ref{Jreal}) for the generators associated to the 
simple roots holds {\em for all real roots $\alpha$}. This can be seen by decomposing 
a given (positive) real root $\alpha$ into two other (positive) real roots $\beta$ 
and $\gamma$ by $\alpha=\beta+\gamma$. Whenever this is possible, formula 
(\ref{ABcomm}) implies that $\alpha$ is represented by (\ref{Jreal}) if $\beta$ and 
$\gamma$ are. Since $E_{10}$ has only single edges in its Dynkin diagram, 
such a decomposition appears always possible.\footnote{We have verified this 
for all real roots up to height $100$. It is crucial here that $E_{10}$ is `simply-laced' in the sense that it has only single lines in its Dynkin diagram. For finite-dimensional simple Lie algebras this requirement is equivalent to having a symmetric Cartan matrix. However, in the Kac--Moody case these notions are no longer equivalent and the relevant property here is having only single edges (corresponding to having only values $-1$ or $0$ off the diagonal of the Cartan matrix). This entails that there is a single orbit of real roots. For other symmetric Cartan matrices with value $-2$ or smaller appearing (like $AE_3$~\cite{FeFr83}) one has multiple orbits and one can easily construct counterexamples to the decompostion $\alpha=\beta+\gamma$ of positive real roots into sums of positive real roots.} In other words, we have found an explicit representation
of the $K(E_{10})$ generators for an infinite number of real roots. For any such
root $\alpha$ we can explicitly exponentiate  the action of the associated 
$K(E_{10})$ element. To this end, we define the projection  operators 
\begin{equation}
\Pi_1(\alpha)_{\ta\tb} = \frac12 \alpha_\ta \alpha_\tb \;\; , \quad
\Pi_2(\alpha)_{\ta\tb} = G_{\ta\tb} - \frac12 \alpha_\ta \alpha_\tb 
\end{equation}
such that (for $i,j=1,2$)
\begin{equation}
\Pi_i \Pi_j = \delta_{ij} \Pi_j \; , \quad \Pi_1 + \Pi_2 = \id 
\end{equation}
and
\begin{equation}
X(\alpha)_{\ta\tb} = \frac14 \Pi_1(\alpha)_{\ta\tb} - \frac34 \Pi_2(\alpha)_{\ta\tb}
\end{equation}
It is then straightforward to show that\footnote{This factorised form of the action of $K(E_{10})$ was crucial in the investigation of the fermionic billiard in~\cite{Damour:2009zc}.}
\begin{eqnarray}
e^{\omega \hJ(\alpha)} \qphi^\ta e^{-\omega \hJ(\alpha)} &=&
\left[ \cos \frac{\omega}2 - \sin\frac{\omega}2 \Gamma(\alpha) \right] 
\Pi_1(\alpha)_{\ta\tb} \qphi^\tb \nn\\[2mm]
&+ & \left[ \cos \frac{3\omega}2 + \sin\frac{3\omega}2 \Gamma(\alpha) \right] 
\Pi_2(\alpha)_{\ta\tb} \qphi^\tb
\end{eqnarray}
In particular, we see that for a rotation about $\omega=2\pi$, the operator $\qphi^\ta$
is mapped to $-\qphi^\ta$, as it should be for a fermion.

While the simple form (\ref{Jreal}) is valid for real roots  it no longer holds for 
imaginary roots. Nevertheless, given that the generating relations (\ref{ke10relations})
are obeyed, the general formula (\ref{ABcomm}) fixes the action of $K(E_{10})$ 
{\em for arbitrary generators}.  One new feature for imaginary roots is that the root spaces can be degenerate, that is, there are several independent $K(E_{10})$ generators for a (positive) root $\alpha$.  Formula (\ref{ABcomm}) implies that the spinor part  is given by $\Gamma$-matrices parametrised by the root vector $\alpha$ as in (\ref{GAdef}) but we have to introduce an 
additional label for the number of independent generators in a given root space. 
We generally write
\begin{align}
\hJ^r(\alpha) = \left\{ \begin{array}{rl} Z^r(\alpha)_{(\cA\cB)} \,
\qphi^\cA\Gamma(\alpha)\qphi^\cB,&\text{if $\alpha\cdot\alpha\in4\ints+2$}\\[2mm]
Z^r(\alpha)_{[\cA\cB]} \,
\qphi^\cA\Gamma(\alpha)\qphi^\cB,&\text{if $\alpha\cdot\alpha\in4\ints$}\end{array}\right.
\end{align}
where we have distinguished the symmetries of the coefficient matrix induced by the symmetries of the $\Gamma$ matrix, and where the indices $r,s,\dots$ label 
the multiplicity of the $\alpha$ root space $\mf{g}_\alpha$ , {\it viz.}
\begin{equation}
r,s,\dots = 1,\dots, {\rm dim}\, \mf{g}_\alpha
\end{equation}
Of course, there will still be degeneracies here as well, because the representation is
unfaithful, but they will be less severe than for the Dirac spinor, crudely speaking because
the vector spinor representation is more faithful than the Dirac representation in that the quotient algebra is larger.

So far, we have not found a general formula for $\hJ^r(\alpha)$ for arbitrary $\alpha$,
but these operators are nevertheless implicitly determined by (\ref{ABcomm}), and this 
gives a very interesting new view on the nature of root multiplicities in hyperbolic algebras. 
Let us first consider the affine null root $\delta^2=0$ of $E_{10}$. This root (like
all null roots) is known to have 
multiplicity $8$ and can all be obtained as a sum of real roots~\cite{Kac}. 
Therefore suppose $\delta=\beta^r+\gamma^r$ with $\beta^r$ and $\gamma^r$ 
positive real roots where we have included the label $r$ to indicate that the 
decomposition of $\delta$ is not unique. Using (\ref{Jreal}) and applying 
(\ref{ABcomm}) one finds
\begin{align}
\lb \hJ(\beta^r), \hJ(\gamma^r) \rb = 
Z^r(\delta)_{\ta\tb} \, \qphi^\ta\Gamma(\delta)\qphi^\tb
\end{align}
with 
\begin{align}
 Z^r(\delta)_{\ta\tb}  = -\frac12 \beta^r_{[\ta} \gamma^r_{\tb]} 
 \end{align}
 Note that now the matrix $\Gamma(\delta)$ is symmetric, whereas the associated
 matrix $Z^r$ is anti-symmetric.
For any null root $\alpha$ there are many such decompositions $\alpha=\beta^r+\gamma^r$ but there can be at most $\text{mult}(\alpha)=8$ different generators obtained in this process.
A basic consistency check of this calculation is then that out of the $120$ decompositions 
of the first null root $\delta$ into pairs $(\beta^r,\gamma^r)$ of real roots only eight
combinations $\beta^r_{[\ta}\gamma^r_{\tb]}$ are linearly independent, reproducing 
exactly the degeneracy of the null root. 

Delving deeper into the light-cone it is clear that more general combinations of the 
tensors $Z_{\ta\tb}$ are generated, although no `nice' structure seems to emerge.
However, by counting the number of independent 
bilinears (\ref{VSbils}) we know that the vector spinor realization of $K(E_{10})$
contains at most $\frac12\times 320 \times 319 = 50\, 040$ independent elements. This is the maximum number because $K(E_{10})$ is represented by $320\times 320$ matrices that preserve the bilinear form (\ref{bilvs}). The bilinear form is of signature $\lp (-)^{288}, (+)^{32}\rp$ and therefore the quotient algebra in the case of the vector spinor is a subalgebra of the Lie algebra $\mf{so}(288,32)$. A computer search indicates that all elements can be generated by successive commutation.\footnote{We note that in~\cite{Damour:2013eua} the vector spinor representation of the hyperbolic algebra $AE_3$ (hyperbolic over-extension of $\mf{sl}(2,\reals)$) was studied and it was found there that a (chirality) operator $G_{\ta\tb}\qphi^\ta \gamma^5 \qphi^\tb$ exists that commutes with the action of $K(AE_3)$. In that case this reduces $\mf{so}(8,4)$ to $\mf{u}(4,2)$. A similar operator does not exist for $K(E_{10})$.}
We stress the remarkable fact that the non-simple `compact' subalgebra $K(E_{10})$ 
with negative definite invariant bilinear form has unfaithful representations that 
give rise to \emp{non-compact} quotients!

\section{Higher spin realizations}
\label{sec:HS}

We now use our formalism  to explore new territory. Supergravity is known to stop at spin-$\frac32$, 
and this fact is correlated with the maximum number of $D=11$ dimensions
for maximal supersymmetry and supergravity. However, the expressions that we have 
derived for the vector spinor strongly suggest that there should exist a generalization of
these formulas to higher spin fermions. We therefore proceed by trial and error,
and thus simply postulate the existence of suitable `higher spin' fermionic operators.

The general strategy here 
will be to verify the generating relations (\ref{ke10J})
by searching for novel realizations of the `master identity'
\begin{align}\label{Master1}
\big\{ X(\alpha) , X(\beta)  \big\}_{\cA\cB}  &= \frac12 X_{\cA\cB}(\alpha \pm \beta) 
&&    \textrm{if $\alpha\cdot\beta = \mp 1$}  \nn \\[2mm]
\big[ X(\alpha) , X(\beta)   \big]_{\cA\cB}  &= 0   &&  \textrm{if $\alpha\cdot\beta = 0 $}   
 \end{align}
for different choices of $\cA$ taken among irreducible tensor representations of $S0(1,9)$. 
We emphasize again the unusual feature that the notion of `higher spin' here 
does not refer to ordinary space-time!

\subsection{Spin-$\frac52$}

We start with $\cA=(\ta\tb)$ and an associated operator $\qphi^{\ta\tb}_\alpha$ symmetric in its two upper two indices  $\qphi^{\ta\tb}_\alpha = \qphi^{\tb\ta}_\alpha$. The operator is taken to obey the canonical anticommutation relations
\begin{align}
\label{qb52}
\{ \qphi^{\ta\tb}_\alpha \, , \, \qphi^{\tc\td}_\beta \} = G^{\ta(\tc} G^{\td)\tb} \delta_{\alpha\beta}
\end{align}
In our figurative way of speaking, we will refer to this as a `spin-$\frac52$' fermion.
Because the fermionic spinor $\qphi^{\ta\tb}$ has $32 \times 55$ real components,
the dimension of the associated fermionic Fock space is now $2^{880}$, already
quite large!

We are thus looking for an operator realization
\begin{equation}\label{X52}
\hJ(\alpha) = X(\alpha)_{\ta\tb\,\tc\td}\,\qphi^{\ta\tb} \Gamma(\alpha) \qphi^{\tc\td}
\end{equation}
satisfying the basic relations (\ref{ke10J}). As we explained, this is ensured if
we can find a solution of (\ref{Master1}). Happily, such a solution exists:
For any real root $\alpha$ the associated $X(\alpha)$ is given by\footnote{This solution is the unique non-trivial solution with these three terms. By considering trace terms in the ansatz one can find more solutions, see~(\ref{X52tr}) below.}
\begin{equation}
\label{52X}
X(\alpha)_{\ta\tb\,\tc\td} = \frac12 \alpha_\ta\alpha_\tb \alpha_\tc\alpha_\td
          - \alpha_{(\ta} G_{\tb)(\tc} \alpha_{\td)} + \frac14 G_{\ta(\tc} G_{\td)\tb}
\end{equation}
It is a gratifying little calculation to check that the master identity (\ref{Master}) in the
generalized form (\ref{Master1}) is indeed satisfied. Therefore, 
the generating relation (\ref{ke10relations}) are also satisfied and we have 
thus found a new (still unfaithful) representation of the $K(E_{10})$ algebra.  
This representation is less unfaithful than the vector spinor representation obtained 
from supergravity.  We also give the transformation rules in `first-quantized' form 
acting on a single classical field $\phi^{\ta\tb}$. They are, for real roots $\alpha$,
\begin{align}
J(\alpha) \cdot \qphi^{\ta\tb} \equiv \big[ \hJ(\alpha) \,,\, \qphi^{\ta\tb} \big] 
= -2 X_{\ta\tb\,\tc\td} \Gamma(\alpha) \qphi^{\tc\td}.
\end{align}
The representation space is of dimension $55\times 32= 1760$.

We can now repeat the calculations of the previous section. In particular,
there exists a decomposition of unity in terms of three projection operators;
these are given by
\begin{eqnarray}
\Pi_1(\alpha)_{\ta\tb\,\tc\td} &=& \frac14\alpha_\ta\alpha_\tb \alpha_\tc\alpha_\td \nn\\
\Pi_2(\alpha)_{\ta\tb\,\tc\td} &=& - \frac12 \alpha_\ta\alpha_\tb \alpha_\tc\alpha_\td 
               + \alpha_{(\ta} G_{\tb)(\tc} \alpha_{\td)}         \nn\\
\Pi_3(\alpha)_{\ta\tb\,\tc\td} &=& +\frac14 \alpha_\ta\alpha_\tb \alpha_\tc\alpha_\td 
      - \alpha_{(\ta} G_{\tb)(\tc} \alpha_{\td)} +  G_{\ta(\tc} G_{\td)\tb}          
\end{eqnarray}
obeying (now with $i,j=1,2,3$)
\begin{equation}
\Pi_i \Pi_j = \delta_{ij} \Pi_j \; , \quad \Pi_1 + \Pi_2 + \Pi_3 = \id
\end{equation}
Furthermore
\begin{equation}
X(\alpha) = \frac14 \Pi_1(\alpha) - \frac34 \Pi_2(\alpha) + \frac14 \Pi_3(\alpha)
\end{equation}
Again we can compute the rotation about an angle $\omega$
\begin{eqnarray}
\exp\big(\omega \hJ(\alpha)\big) \qphi^{\ta\tb} \exp\big(-\omega \hJ(\alpha)\big) 
&=&  \left( \cos\frac{\omega}2 - \sin \frac{\omega}2 \,\Gamma(\alpha)\right) 
    \Pi_1(\alpha)^{\ta\tb\,\tc\td} \qphi_{\tc\td}  \nn\\[1mm]
&& \!\!\!\!\!\!\!\!\!\!\!\!\!\!\!\!\!\!\!\!\!\!\!\!\!\!\!\!\!\!\!\!\!\!\!\!
+  \; \left( \cos\frac{3\omega}2 + \sin \frac{3\omega}2 \, \Gamma(\alpha)\right) 
    \Pi_2(\alpha)^{\ta\tb\,\tc\td} \qphi_{\tc\td}  \nn\\[1mm]
&& \!\!\!\!\!\!\!\!\!\!\!\!\!\!\!\!\!\!\!\!\!\!\!\!\!\!\!\!\!\!\!\!\!\!\!\!
+   \; \left( \cos\frac{\omega}2 - \sin \frac{\omega}2 \, \Gamma(\alpha)\right) 
    \Pi_3(\alpha)^{\ta\tb\,\tc\td} \qphi_{\tc\td}      
\end{eqnarray}
In particular this implies, as before,
\begin{equation}
e^{2\pi  \hJ(\alpha)} \qphi^{\ta\tb} e^{-2\pi \hJ(\alpha)} = - \qphi^{\ta\tb}
\end{equation}

{\bf Properties and reducibility of the representation.} 
The representation $\qphi^{\ta\tb}$ is not irreducible. It is possible to 
remove a trace over the $SO(1,9)$ tensor indices: one can verify that
\begin{align}
\qphi = G_{\ta\tb} \qphi^{\ta\tb}
\end{align}
transforms into itself under the action of $K(E_{10})$. Similarly, the traceless 
part $\qphi^{\ta\tb}-\frac1{10}G^{\ta\tb}\qphi$ transforms into itself, and therefore the $1760$ 
components of $\qphi^{\ta\tb}$ decompose into two irreducible representations, 
of dimension $1728$ and $32$, respectively. The tensor (\ref{X52}) is thus 
replaced by
\begin{equation}\label{X52tr}
\tilde X(\alpha)_{\ta\tb\,\tc\td} = \frac12 \alpha_\ta\alpha_\tb \alpha_\tc\alpha_\td
          - \alpha_{(\ta} G_{\tb)(\tc} \alpha_{\td)} + 
          \left[\frac14 G_{\ta(\tc} G_{\td)\tb} - \frac1{40} G_{\ta\tb} G_{\tc\td}\right]
\end{equation}
such that $G^{\ta\tb} \tilde X_{\ta\tb\,\tc\td} (\alpha)= 0$.
We have not determined the quotient algebras. However, a computer investigation indicates that this representation is \emp{more faithful} than the spin-$\frac32$ vector spinor in that there are more independent elements in the quotient than in $\mf{so}(288,32)$. 

The consistency of the representation (\ref{X52}) with (\ref{52X}) implies that there is an invariant bilinear form on the (first quantized) representation space $\phi^{\ta\tb}$ furnished by the canonical bracket (\ref{qb52}):
\begin{align}
\lp \phi | \phi \rp_{\text{spin-5/2}} = \phi^{\ta\tb} G_{\ta\tc} G_{\tb\td} \phi^{\tc\td}.
\end{align}
The eigenvalues of this invariant bilinear form are $(-18)^1$, $(-8)^9$ and $(2)^{45}$, where the exponents designate the degeneracy of a given eigenvalue. The single most negative eigenvalue $-18$ is associated with the Dirac spinor trace $G_{\ta\tb}\phi^{\ta\tb}$, so that on the irreducible traceless part one has an invariant form of signature $(-)^9,(+)^{45}$, suggesting that the quotient in this case $\mf{so}(1440,288)$.

{\bf $SO(10)$-interpretation of the representation.} 
Because $K(E_{10})$ has a natural subalgebra $\mf{so}(10)$ generated by 
the $x_i$ for $i=1,\ldots,9$ (cf. figure~\ref{fig:E10}), any representation space 
of $K(E_{10})$ must be a representation of $\mf{so}(10)$ as well. Hence, the 
finite-dimensional `spin-$\frac52$' representation $\qphi^{\ta\tb}$ must be completely 
reducible under the action of $\mf{so}(10)$. However, it is also clear that the 
transcription between the $SO(1,9)$ form and the $SO(10)$ form of the 
generators can no longer be achieved by means of the simple formula (\ref{newvs})
(for instance, acting with two $\Gamma$ matrices does not give anything useful
because the $\Gamma$-matrices are contracted with a symmetric tensor).

One can nevertheless analyse the $SO(10)$ representation content by 
determining the weight diagram under the $\mf{so}(10)$ Cartan generators.
In this way one finds the weight diagram of a reducible $\mf{so}(10)$ representation of type
\begin{align}\label{1760}
{\bf 1760} \to {\bf 1120} \oplus 2\times{\bf 288} \oplus 2\times {\bf 32}.
\end{align}
Writing these representations as (reducible) tensor-spinors of $\mf{so}(10)$ yields two fields
\begin{align}
\psi^{ab}=\psi^{[ab]}\quad\text{and}\quad \psi^a,
\end{align}
where both have all possible $\Gamma$-traces. Whereas the original construction in 
mini-superspace was based on an operator $\qphi^{\ta\tb}$ that was \emp{symmetric} 
in its $SO(1,9)$ indices, the same $K(E_{10})$ representation is \emp{anti-symmetric} 
when viewed from the spatial rotation group $SO(10)$.\footnote{This means that it could be interpreted as the curl $\partial_{[a}\psi_{b]}$ of the standard gravitino in a fermionic extension of the gradient hypothesis~\cite{Damour:2002cu}.} However, this seemingly strange 
feature is fully explained by considering the decomposition under the $SO(16)$ subgroup
of $K(E_{10})$, cf. section~\ref{sec:Truncations} below.

\subsection{Spin-$\frac72$}

Next we consider `spin-$\frac72$' and the associated fermion operator
 $\qphi^{\ta\tb\tc}_\alpha=\qphi^{(\ta\tb\tc)}_\alpha$, further generalising (\ref{X52}) to
\begin{equation}\label{X72}
\hJ(\alpha) = X(\alpha)_{\ta\tb\tc\,\td\te\tf}\,
\qphi^{\ta\tb\tc} \Gamma(\alpha) \qphi^{\td\te\tf}.
\end{equation} 
The canonical commutation relations for the operator $\qphi^{\ta\tb\tc}$ are taken as
\begin{align}\label{CB72}
\ld \qphi^{\ta\tb\tc}_\alpha,\qphi^{\td\te\tf}_\beta\rd = G^{\ta\td}G^{\tb\te}G^{\tc\tf}\delta_{\alpha\beta},
\end{align}
where the right-hand-side is assumed symmetrised over $(\ta\tb\tc)$ and $(\td\te\tf)$ according to the symmetries of $\qphi^{\ta\tb\tc}$.

It is now more tedious to find a solution of the master identity~(\ref{Master1}).
By employing a computer program, we have checked that it is now solved by the tensor
\begin{align}
\label{X72}
X_{\ta\tb\tc}{}^{\td\te\tf}(\alpha)& =-\frac13 \alpha_\ta \alpha_\tb\alpha_\tc \alpha^\td\alpha^\te\alpha^\tf
+\frac32 \alpha_{(\ta} \alpha_{\tb}\delta_{\tc)}^{(\td} \alpha^{\te}\alpha^{\tf)}
-\frac32\alpha_{(\ta}\delta_{\tb}^{(\td} \delta_{\tc)}^{\te\phantom{)}}\alpha^{\tf)}_{\phantom{)}}&\nn\\
&\quad +\frac14 \delta_{(\ta}^{(\td}\delta_{\tb\phantom{)}}^{\te\phantom{)}}\delta_{\tc)}^{\tf)}
+\frac1{12}(2-\sqrt{3}) \alpha_{(\ta}  G_{\tb\tc)}G^{(\td\te}\alpha^{\tf)}&\\
&\quad+\frac1{12}(-1+\sqrt{3})\lp \alpha_{(\ta}\alpha_{\tb}\alpha_{\tc)}G^{(\td\te}\alpha^{\tf)}
+\alpha_{(\ta}  G_{\tb\tc)}\alpha^{(\td}\alpha^\te\alpha^{\tf)}\rp.\nn
\end{align}
Here, we have raised the second set of indices $(\td\te\tf)$ with the Lorentz metric $G^{\ta\tb}$ in order to make the formula easier to read. The dimension of this representation (in first quantized form) is $32\times 220=7040$.

There is an invariant bilinear form on this representation furnished by
\begin{align}\label{bil72}
\lp \phi| \phi\rp_{\text{spin-7/2}} = \phi^{\ta\tb\tc}G_{\ta\td}G_{\tb\te}G_{\tc\tf} \phi^{\td\te\tf}.
\end{align}

The result (\ref{X72}) was found to be the essentially unique non-trivial solution 
with the most general ansatz for $X_{\ta\tb\tc}{}^{\td\te\tf}$ that is made out of 
components $\alpha_\ta$ and $\delta_\ta^\tb$. Note also that, unlike for spin-$\frac52$,
we can{\em not} remove a trace. We should nevertheless point out that there
may be other solutions if the above ansatz can be suitably relaxed. Indeed, the assumed forms of
the anticommutation relations (\ref{CB72}) and the bilinear invariant (\ref{bil72}) could
conceivably be modified by other terms preserving $SO(1,9)$ invariance. Furthermore,
the calculation could be altered by assuming admixtures of lower spin. We will,
however, leave to future work the further exploration of these possibilities and the search
for a systematic pattern underlying the present construction that might pave the way to
fermionic representations of yet higher spin. 

Finally one can try a similar ansatz for spin-$\frac92$. A computer search using the Gamma~\cite{Gran:2001yh} and xAct~\cite{xact} packages then 
shows that with the most general ansatz (including trace terms) there do not appear
to exist any non-trivial solutions to the master identity  (\ref{Master1}). Below we explain 
why this negative outcome is, in fact, perfectly consistent within our general framework.

\section{Truncations}
\label{sec:Truncations}

Because our results are valid for $E_{10}$ and its compact subgroup $K(E_{10})$
they must {\em a fortiori} also hold for the truncation to their affine and finite 
subgroups via the chain of embeddings
\begin{equation}
E_8\subset E_9\subset E_{10} \quad \Longleftrightarrow\quad
K(E_8)\subset K(E_9) \subset K(E_{10})
\end{equation}
As we will see, this leads to important restrictions on the possible fermionic  higher
spin representations that will, in particular, confirm the above findings 
concerning spin-$\frac52$, spin-$\frac72$ and  spin-$\frac92$.

\subsection{$E_8$ and $K(E_8)\equiv Spin(16)$}

Truncating to the finite dimensional subgroups $E_8$ and $K(E_8)$ 
it is obvious that our fermion operators must belong to representations
of the group $K(E_8)\equiv Spin(16)$. To study  this truncation we discard the
simple roots $\alpha_1$ and $\alpha_2$ from the list (\ref{wallbs}). The matrix
$\Gamma(\alpha_1)\equiv \Gamma_{12}$ associated to the root $\alpha_1$ can
then be used to decompose the 32-component Dirac spinor of $SO(10)$ into
 a pair of spinors, each one of which has only 16 (real) components.
Likewise the indices $\ta,\tb,\dots$ now only run from 3 to 10 (or, more conveniently,
from 1 to 8).  For the Dirac and vector spinor representations we thus arrive at the 
following (rather obvious) association
\begin{subequations}
\begin{align}\label{1232}
&\chi_\alpha \quad &\leftrightarrow  &\quad {\bf 16}_v &\quad & \leftrightarrow  \quad \varphi^I 
\\[1mm]
&\phi^\ta_\alpha  \quad &\leftrightarrow &\quad {\bf 128}_s  &\quad & \leftrightarrow \quad  \chi^A 
\end{align}
\end{subequations}
where the $Spin(16)$ representations appear in the middle column (with the $Spin(16)$ 
vector and spinor indices $I,J,\dots$ and $A,B,\dots$, respectively, in the last column). 
Counting components for spin-$\frac52$ and spin-$\frac72$ we see that
\begin{subequations}
\begin{align}\label{5272}
&\tilde\phi^{\ta\tb}_\alpha \quad &\leftrightarrow  &\quad {\bf 560}_v &\quad 
& \leftrightarrow  \quad \varphi^{IJK} \\[1mm]
&\phi^{\ta\tb\tc}_\alpha  \quad &\leftrightarrow &\quad {\bf 1920}_s  
&\quad & \leftrightarrow \quad  \chi^{IA} 
\end{align}
\end{subequations}
where $\varphi^{IJK}\equiv \varphi^{[IJK]}$ is fully antisymmetric, and $\chi^{IA}$
denotes the traceless vector spinor of $Spin(16)$, with $\Gamma_{\dot A A}\chi^{IA} =0$.
The tilde on $\tilde{\phi}^{\ta\tb}$ also indicates tracelessness: $G_{\ta\tb} \tilde\phi^{\ta\tb} = 0$, 
whereas no trace can be taken out of $\phi^{\ta\tb\tc}$. We have also verified that the identification (\ref{5272}) is 
correct by computing its full $\mf{so}(16)$ weight diagram. As required by 
consistency, the decomposition of the ${\bf 560}_v$ under the 
$SO(8)$ subgroup of $SO(10)$ in (\ref{1760}) matches with its decomposition 
under the diagonal $SO(8)=[SO(8)\times SO(8)]_{\rm diag}$ subgroup of $SO(16)$.
Equally important, the fact that the $SO(16)$ representations in (\ref{1232}) 
and (\ref{5272}) are different, demonstrates that the spin-$\frac52$ and 
spin-$\frac72$ representations of $K(E_{10})$ are genuinely different from the 
Dirac and the vector spinor representations.

By contrast, if we repeat the counting for the hypothetical spin-$\frac92$ realization
with fermionic operator $\phi^{\ta\tb\tc\td}$  we get $330\times 16$ components. 
There is no irreducible representation of $Spin(16)$ of that dimension
%~\footnote{Barring the trivial possibility of simply taking an appropriate multiple 
%of the ${\bf 16}_v$ representation, which we exclude here.} 
(this conclusion remains unaltered if we take  out the trace w.r.t. $G_{\ta\tb}$). 
For this reason such a realization cannot exist, explaining the failure of our 
computer search. Conversely, one can use the condition that there {\em must} exist an
associated representation of $Spin(16)$ as a guiding principle towards a systematic
search for fermionic realizations of yet higher spin and towards their explicit construction 
(where also non-trivial Young tableaux for the $SO(1,9)$ indices $\ta,\tb,\dots$ will
have to be considered). Preliminary searches indicate that the condition of compatibility with $Spin(16)$ is rather restrictive.

\subsection{$E_9$ and $K(E_9)$}

As a further special case we can apply the above formulas to the affine subgroup
and compare to the results of \cite{KNP}. This is done by specializing the hyperbolic
roots to affine real roots, which are generally of the form
\begin{equation}
\label{E9rt}
\alpha = m\delta + \alpha'
\end{equation}
where $\delta$ is the affine null root of $E_9$, and $\alpha'$ any root of $E_8$.
Because $\delta = (0,1,1,1,1,1,1,1,1,1)$ in the basis (\ref{wallbs}), the associated
$\Gamma$ matrix is
\begin{equation}
\Gamma(\delta) = \Gamma^{2\,3\,4\,5\,6\,7\,8\,9\,10} =  \Gamma^{0} \Gamma^{1}
\equiv -\Gamma^*
\end{equation}
where the last matrix is just the $\gamma^5$ matrix in two space-time 
dimensions as identified in \cite{KNP}. We also require
\begin{equation}
\Gamma(\alpha) = \eps_{\alpha',m\delta} \Gamma(\alpha') \Gamma(m\delta)
\end{equation}
with $\Gamma(m\delta)=(-\Gamma^*)^m$ and $\eps_{\alpha',m\delta} =
(\eps_{\alpha',\delta})^m$, where $\eps_{\alpha',\delta} =
 (-1)^{\text{ht}(\alpha')}$ in terms of the height of an $E_8$ root $\alpha'$.

We can then recover the expressions given in equation (3.10) of~\cite{KNP} for both 
the Dirac representation and the vector spinor representation; for the former we get
\begin{equation}
\hJ(\alpha) = \frac14 \eps_{\alpha',m\delta} \,\qchi \Gamma(\alpha') (-\Gamma^*)^m\qchi.
\end{equation}
while for the vector spinor, plugging (\ref{E9rt}) into (\ref{X32}), one finds
\begin{equation}
\hJ(m\delta + \alpha') = \eps_{\alpha',m\delta}\lp-\frac12(m\delta +\alpha')_\ta (m\delta + \alpha')_\tb+\frac14G_{\ta\tb}\rp
 \qphi^\ta \Gamma(\alpha') (-\Gamma^*)^m\qphi^\tb
\end{equation}
thus recovering the chiral nature of the transformations and the quadratic dependence 
on the affine level $m$ already exhibited in equations (3.23) and (3.27) of~\cite{KNP}, 
see also~\cite{NS}. The expressions above are valid for real roots.

Continuing to the new representations of `spin-$\frac52$' and `spin-$\frac72$', one needs to plug (\ref{E9rt}) into (\ref{52X}) and (\ref{X72}).
The salient feature of the new expressions is that they now 
become \emp{quartic} and \emp{sextic} in $m$, respectively. In terms of a current algebra realization of $K(E_{9})$ as studied in~\cite{NS,KNP} this implies that the representations 
now depend on derivatives w.r.t. the spectral parameter up to order $4$ or $6$!

\section{Weyl group}
\label{sec:WG}

The Weyl group of $E_{10}$ can be represented in any integrable module as being generated by the ten fundamental reflections~\cite{Kac}
\begin{align}
w_i^{E_{10}} = e^{f_i}e^{-e_i}e^{f_i}
\end{align}
for $i=1,\ldots,10$. This is a well-defined operator in an integrable module for $E_{10}$ since there , by definition, the simple Chevalley generators are nilpotent, rendering the exponentials well-defined. In fact, any such integrable module branches into an (infinite) sum of finite-dimensional $\mf{sl}(2,\reals)$ modules for each of which one can verify that
\begin{align}
\label{wKE}
w_i:= w_i^{E_{10}} = e^{\frac{\pi}2 x_i}
\end{align}
with $x_i=e_i-f_i$ as before. The formulas give an extension of the $E_{10}$ Weyl group by an abelian group $\tilde{D}$ (generated by the squares $w_i^2$) that is a normal subgroup of the (discrete) group of transformations generated by the $w_i^{E_{10}}$. The quotient of the group $\tilde{W}$ generated by the $w_i$ in an $E_{10}$ representation by the normal subgroup $\tilde{D}$ is isomorphic to the Weyl group $W(E_{10})$~\cite{Kac}.

Since (\ref{wKE}) is written solely in terms of the $K(E_{10})$ generator $x_i$ it is 
possible to also define the action of the fundamental on a space that is only a 
representation of $K(E_{10})$ but not of $E_{10}$, as is the case for the unfaithful spinors considered in the present paper. As was shown in~\cite{Damour:2009zc,Koehl} this 
leads to a presentation of the Weyl group of $E_{10}$ in a more generalized form: 
The group $\tilde{W}_{\text{spin}}$ generated by the $w_i$ of (\ref{wKE}) has a normal 
non-abelian subgroup $\tilde{D}_{\text{spin}}$ (generated by the $w_i^2$). 
The quotient \emp{can} be isomorphic to $W(E_{10})$, as is the case for example for 
the vector spinor, but for the Dirac spinor it is actually a finite group. A notable feature 
of the group $\tilde{W}_{\text{spin}}$ is that the order of the fundamental reflection 
increases: One has $w_i^8=\id$ whereas in the known (bosonic) representation 
of $E_{10}$ one has $w_i^4=\id$. This increase in the order is in agreement with 
the interpretation of the representation as fermions that have to be rotated twice to 
come back to themselves.

We now analyse the group $\tilde{W}_{\text{spin}}$ in our formulation and for the new representations, defining more generally the reflection in  a real root $\alpha$ 
by conjugation with the operator
\begin{equation}
w_\alpha %= \exp\left( \frac{\pi}2 (E(\alpha) - E(-\alpha) \right) 
\equiv \exp\left( \frac{\pi}2 J(\alpha)\right).
\end{equation}
The normal subgroup is generated by the squares 
$w_\alpha^2=\exp\left( \pi J(\alpha)\right)$.
To compute the explicit representatives it is simplest to decompose the tensors
$X_{\cA\cB}(\alpha)$ in terms of projection operators. In this way we get
\begin{subequations}
\begin{align}
w_\alpha &= \frac1{\sqrt{2}}\big(\id - \Gamma(\alpha)\big)     && \text{for Dirac}\\
w_\alpha &= \frac1{\sqrt{2}}\big(\id - \Gamma(\alpha)\big) 
   \big[\Pi_1(\alpha) - \Pi_2(\alpha)\big]
   && \text{for the vector spinor}\\
w_\alpha &= \frac1{\sqrt{2}}\big(\id - \Gamma(\alpha)\big)
 \big[\Pi_1(\alpha) - \Pi_2(\alpha) + \Pi_3(\alpha)\big] && \mbox{for spin-$\frac52$}
\end{align}
\end{subequations}
These expressions are consistent with the general form %for {\em any} spin
\begin{equation}
w_\alpha =  \frac1{\sqrt{2}}\big(\id - \Gamma(\alpha)\big)   \sum_j \varepsilon_j \Pi_j
\end{equation}
with $\varepsilon_j = \pm 1$ and suitable projectors $\Pi_j$. In this form it is
obvious that
\begin{align}
w_\alpha^2 = -\Gamma(\alpha)
\end{align}
because of the orthogonality and completeness of the projectors. Therefore the 
subgroup $\tilde{D}_{\text{spin}}$ generated by the squares is the \emp{Clifford group} 
of all gamma matrices (with sign). This is a non-abelian group of order $2048$ 
(see also~\cite{Damour:2009zc}), which is obviously a normal subgroup.

The Coxeter relations can be verified in the quotient $\tilde{W}_{\text{spin}}/\tilde{D}_{\text{spin}}$ as follows. The only relations one needs to check are
\begin{subequations}
\begin{align}
(w_\alpha w_\beta)^2 &= \id\quad\Leftrightarrow&\quad w_\alpha w_\beta&= w_\beta w_\alpha&&\text{for $\alpha\cdot\beta=0$},\\
(w_\alpha w_\beta)^3 &= \id\quad\Leftrightarrow&\quad w_\alpha w_\beta w_\alpha&= w_\beta w_\alpha w_\beta&&\text{for $\alpha\cdot\beta=\pm 1$}.
\end{align}
\end{subequations}
To check them we do not even need to make recourse to the explicit representations but can verify them directly from the commutation relations (\ref{ke10J}) of $K(E_{10})$. One has for $\alpha\cdot\beta=0$
\begin{align}
e^{\frac{\pi}2 J(\alpha) } J(\beta) e^{-\frac{\pi}2 J(\alpha) } = J(\beta)
\end{align}
since $J(\alpha)$ and $J(\beta)$ commute. Exponentiating this relation and bringing the right-most factor to the other side yields the wanted
\begin{align}
e^{\frac{\pi}2 J(\alpha) }e^{\frac{\pi}2 J(\beta) } =e^{\frac{\pi}2 J(\beta) }e^{\frac{\pi}2 J(\alpha) }
\quad\Leftrightarrow \quad
w_{\alpha} w_{\beta} = w_{\beta} w_{\beta}.
\end{align}
Similarly, one has for $\alpha\cdot \beta=\mp 1$ that
\begin{align}
e^{\frac{\pi}2 J(\alpha) } J(\beta) e^{-\frac{\pi}2 J(\alpha) } = \eps_{\alpha,\beta} J(\alpha\pm \beta).
\end{align}
Since we are working in the quotient the right-most factor on the left-hand side can also be written with the inverse sign (as $w_\alpha^2=\id$ in the quotient). Exponentiating this relation one finds
\begin{align}
e^{\frac{\pi}2 J(\alpha) }e^{\frac{\pi}2 J(\beta)} e^{\frac{\pi}2 J(\alpha) } = e^{\frac{\pi}2J(\alpha\pm \beta)},
\end{align}
where the sign $\eps_{\alpha,\beta}$ was suppressed since it is irrelevant in the quotient. Clearly, this relation is symmetric under the exchange of $\alpha$ and $\beta$ thus yielding the cubic Coxeter relation.

\vspace{3mm}

{\bf Acknowledgements:}
We are very grateful to Thibault Damour and Philippe Spindel for informative discussions of their work~\cite{Damour:2013eua} and to Alex Feingold for discussions on Weyl groups.

%%%%%%%%%%%%%%%%%%%%%%%

\end{document}